\documentclass[prl,twocolumn,showpacs,floatfix,amsfonts]{revtex4}
\usepackage{graphicx,graphics,color,epsfig}
\usepackage{bm}
\usepackage{amsmath}
\usepackage{amssymb}
\usepackage{fancyheadings}
\begin{document}
\preprint{}
\title{Elasticity-Driven Nanoscale Electronic Structure in Superconductors}

\author{Jian-Xin Zhu, K. H. Ahn, Z. Nussinov, T. Lookman, A. V. Balatsky, and A. R. Bishop}
\affiliation{Theoretical Division, Los Alamos National Laboratory,
Los Alamos, New Mexico 87545}
\date{April 29, 2003}

\begin{abstract}
The effects of long-range anisotropic elastic deformations on
electronic structure in superconductors are analyzed within the
framework of the Bogoliubov-de Gennes equations. Cases of twin
boundaries and isolated defects are considered as illustrations.
We find that the superconducting order parameter is depressed in
the regions where pronounced lattice deformation occurs. The
calculated local density of states suggests that the electronic
structure is strongly modulated in response to lattice
deformations, and propagates to longer distances. In particular,
this allows the trapping of low-lying quasiparticle states around
defects. Some of our predictions can be directly tested by STM
experiments.

\end{abstract}

\maketitle

In many complex electronic materials such as cuprates, manganites,
ferroelastic martensites, and titanates,  unexpected and puzzling
multiscale modulations of charge, spin, polarization, and strain
variables have been revealed by high resolution
microscopy~\cite{Multi1}. The nonuniform textures found in these
doped materials indicate that their origin is intrinsic: they
arise from coupling between various degrees of freedom. The
textures fundamentally affect local and mesoscopic electronic,
magnetic and structural properties, which are central to the
functionality of correlated electronic materials. There is ample
evidence for significant coupling amongst the electronic degrees
of freedom with the lattice distortion in cuprates and manganites.
The charge carrier doping can act as a local stress to deform
surrounding unit cells~\cite{Multi1,Ahn02}. We might employ a
Landau-Ginzburg (LG) theory to study the coupling between the
electronic (Cooper pair) and lattice (strain tensor) degrees of
freedom in superconductors.
 However, the LG theory can only describe the
long wavelength behavior. The local electronic properties and
lattice distortion necessitate a treatment at the atomic scale.
Recently, we have developed an atomic scale theory for determining
lattice distortions by using strain related variables and their
constraint equations~\cite{Ahn02}. This now enables a systematic
study of the influence of strain on electronic wavefunctions. Here
we apply a microscopic theory to study the order parameter and
local quasiparticle properties in both $s$ and $d$-wave
superconductors.

We consider the following model on a
square lattice:
\begin{eqnarray}
\mathcal{H}&=&-\sum_{ij,\sigma} \tilde{t}_{ij}
c_{i\sigma}^{\dagger}c_{j\sigma} +\sum_{i,\sigma}
(\epsilon_{i}-\mu) c_{i\sigma}^{\dagger}c_{i\sigma} \nonumber \\
&&+\sum_{ij} (\Delta_{ij} c_{i\uparrow}^{\dagger}
c_{j\downarrow}^{\dagger} +\Delta_{ij}^{*} c_{j\downarrow} c_{
i\uparrow} ) \;. \label{EQ:MFA}
\end{eqnarray}
Here  $c_{i\sigma}$ annihilates an electron of spin $\sigma$ on
site $i$. The quantities $\epsilon_{i}$ and $\mu$ are the on-site
impurity potential (if any) and the chemical potential,
respectively. The hopping integral  $\tilde{t}_{ij}$ is modified
by the lattice distortion. The electron-lattice coupling is
approximated by $t_{ij}=t_{ij}^{0}[1-\alpha \epsilon_{ij}]$, where
$t_{ij}^{0}$ is the bare hopping integral, $\epsilon_{ij}$ is the
lattice-distortion variable, and $\alpha$ is the coupling
constant. In our nearest neighbor realization, the bare hopping
integral $t_{ij}^{0}$ is $t$ for nearest neighbor sites and zero
otherwise. Specifically, we take the form of the lattice
distortion to be: $\epsilon_{ij}=[\vert
(\mathbf{R}_{j}+\mathbf{d}_{j})-
(\mathbf{R}_{i}+\mathbf{d}_{i})\vert/\vert
\mathbf{R}_{j}-\mathbf{R}_{i}\vert -1]$, where
$\{\mathbf{R}_{i}\}$ are the undistorted lattice coordinates and
$\{\mathbf{d}_{i}\}$ the lattice displacement vectors with respect
to $\{\mathbf{R}_{i}\}$. We assume an effective superconducting
gap function given by $\Delta_{ij}=\frac{U_{ij}}{2}\langle
c_{i\uparrow} c_{j\downarrow}
-c_{i\downarrow}c_{j\uparrow}\rangle$, where $U_{ij}=U\delta_{ij}$
(i.e., attractive Hubbard-$U$ model) for $s$-wave
superconductivity and $U_{ij}=V\delta_{i+\gamma,j}$ (with $\gamma$
specifying the nearest neighbors to the $i$-th site) for $d$-wave
superconductivity.
By performing a Bogoliubov-Valatin
transformation, we may diagonalize our Hamiltonian
by solving the Bogoliubov-de Gennes (BdG)
equation~\cite{deGennes89}:
\begin{equation}
\sum_{j} \left(
\begin{array}{cc}
{\cal H}_{ij} & \Delta_{ij}  \\
\Delta_{ij}^{*} & -{\cal H}_{ij}^{*}
\end{array}
\right) \left(
\begin{array}{c}
u_{j}^{n} \\ v_{j}^{n}
\end{array}
\right) =E_{n} \left(
\begin{array}{c}
u_{i}^{n} \\ v_{i}^{n}
\end{array}
\right)  \;, \label{EQ:BdG}
\end{equation}
subject to the self-consistency conditions for the superconducting
(SC) order parameter (OP):
\begin{equation}
\Delta_{ij}=\frac{U_{ij}}{4}\sum_{n} (u_{i}^{n}v_{j}^{n*}
+v_{i}^{n*}u_{j}^{n} ) \tanh \left(
\frac{E_{n}}{2k_{B}T}\right)\;. \label{EQ:Self-consistency}
\end{equation}
Here the single particle Hamiltonian reads ${\cal
H}_{ij}=-\tilde{t}_{ij} + (\epsilon_{i}-\mu)\delta_{ij}$. The
quasiparticle wavefunction, corresponding to the eigenvalue $E_n$,
consists of electron ($u_{i}^{n}$) and hole ($v_{i}^{n}$)
amplitudes. The quasiparticle energy is measured with respect to
the chemical potential.

We solve the BdG equations self-consistently by starting off with
an initial gap function. After exactly diagonalizing
Eq.~(\ref{EQ:BdG}), the obtained wavefunction is substituted into
Eq.~(\ref{EQ:Self-consistency}) to compute a new gap function. We
then use this as an input to repeat the above procedure until a
desired convergence is achieved. Below, we report our results for
two types of local lattice distortions at zero temperature--- a
superlattice formed by twin boundaries and a single defect. We
measure the length and energy in units of $a_0$ (the undistorted
lattice constant) and $t$. The chemical potential  $\mu=0$ and no
extrinsic impurity scattering is introduced ($\epsilon_{i}=0$).
The pairing interaction for both the $s$-wave ($U$) and $d$-wave
($V$) superconductors is taken to be 3. The typical system size is
$N_{L}=32\times 32$ and periodic boundary conditions are applied.
When the local quasiparticle density of states (LDOS) is computed,
we implement a much larger system using the above small system as
a supercell. In the absence of distortions, Eq.~(\ref{EQ:BdG}) is
readily solved by resorting to translational invariance: the
resulting quasiparticle energy
$E_{\mathbf{k}}=\sqrt{\xi_{\mathbf{k}}^{2}+
\Delta_{\mathbf{k}}^{2}}$ with $\xi_{\mathbf{k}}=-2(\cos k_{x}
+\cos k_{y})-\mu$, and the energy gap
$\Delta_{\mathbf{k}}=\Delta_{s0}$ or
$\Delta_{\mathbf{k}}=\frac{\Delta_{d0}}{2}(\cos k_{x} -\cos
k_{y})$ for the $s$ or $d$-wave superconductor respectively. For
the given parameter values, we obtain $\Delta_{s0}=0.85$ and
$\Delta_{d0}=1.7$. In both cases, the superconducting coherence
length is $ \sim 2a_{0}$. These values are exaggerated when
compared to real materials: The choice of values is motivated by
the desire to enhance the effect of the lattice distortions.

\begin{figure}
\centerline{\psfig{figure=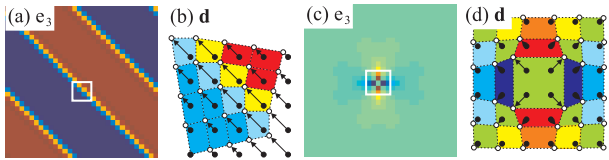,width=8.0cm,angle=0}} \caption{
Strain-$e_{3}$ mode for a periodic twinned microstructure (a) and
a single defect (c) together with their corresponding displacement
configurations [(b) and (d)] within the highlighted window.
$N_{L}=32\times 32$.} \label{FIG:Lattice}
\end{figure}

Before proceeding to the BdG calculation sketched above, we
generate (following~\cite{Ahn02}) two atomic scale lattice
distortions. These distortions arise from {\em long-range
anisotropic} interactions between strains to maintain the
compatibility constraints. The $e_{3}$ (square to rectangle)
strain mode for a periodic twinned microscopic structure and the
corresponding atomic displacements are depicted in
Fig.~\ref{FIG:Lattice}(a-b). The domain consists of rectangular
distortion (red and blue) separated by a domain wall where
$e_{3}=0$. Note that similar but more ``realistic''
microstructures have also been obtained using Monte Carlo
techniques~\cite{Goldman95}. Similar quantities for a single
defect are displayed in panels (c-d). Near the defect, four
alternating distortions are formed in (clover leaf) ``$d$-wave
like'' pattern.

\begin{figure}
\centerline{\psfig{figure=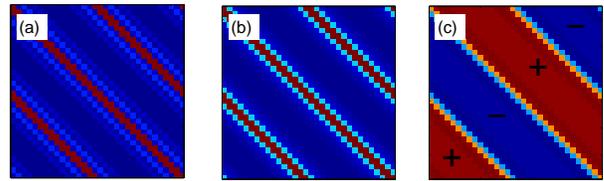,width=8.0cm,angle=0}}
\caption{Spatial variation of the SC OP for periodic twin
boundaries displayed in Fig.~\ref{FIG:Lattice}(a)--- (a) The
$s$-wave OP in an $s$-wave superconductor, and (b) the $d$-wave
and (c) extended $s$-wave components of the OP in a $d$-wave
superconductor. The electron-lattice coupling constant
$\alpha=3$.}
 \label{FIG:OP-TB}
\end{figure}

In Fig.~\ref{FIG:OP-TB}, we show the spatial variation of the SC
OP induced by the deformation of Fig.~\ref{FIG:Lattice}(a) in both
$s$ and $d$-wave superconductors. In both types of
superconductors, the OP is lowered within the domain and is
elevated at the domain wall (Fig.~\ref{FIG:OP-TB}(a-b)).
The magnitude of the OP is depressed in comparison to an
undistorted square lattice since the lattice deformation changes
the band structure, leading to a reduction in  normal density of
states at the Fermi energy. Even at the domain wall, where the
strain induced deformation is weakest, the amplitude of the
enhanced OP is smaller than its value in an undistorted
 square lattice. This is due to the confinement from the
two neighboring domains. In an $s$-wave superconductor, the
relative orbital motion between two paired electrons has an
angular momentum $l=0$, which has the highest symmetry and
generates no subdominant OPs with lower symmetries. However, for a
$d$-wave superconductor, the relative motion between two paired
electrons has the angular momentum $l=2$, which upon scattering
from any inhomogeneity can generate a subdominant OP with symmetry
not lower than that of $l=2$. The $d$-wave OP is defined by
$\Delta_{d}(i)=(\Delta_{i,i+\hat{x}}+\Delta_{i,i-\hat{x}}
-\Delta_{i,i+\hat{y}}-\Delta_{i,i-\hat{y}})/4$, whereas the
extended $s$-wave OP is
$\Delta_{s}(i)=(\Delta_{i,i+\hat{x}}+\Delta_{i,i-\hat{x}}
+\Delta_{i,i+\hat{y}}+\Delta_{i,i-\hat{y}})/4$. In a twinned
domain of a $d$-wave SC, a subdominant extended $s$-wave component
is generated in a real combination $d\pm s$. Because the
symmetries of two twinned domains are reflected into each other
with respect to the twin boundary, the relative phase between the
$d$- and $s$-wave components switches by $\pi$  when a twin
boundary is crossed (Fig.~\ref{FIG:OP-TB}(c)). It has been argued
phenomenologically~\cite{Multi5} that a local
time-reversal-symmetry-breaking state exists at a twin or grain
boundary of YBa$_{2}$Cu$_{3}$O$_{7-\delta}$. Within numerical
accuracy, our result shows a real admixture of the $d$-wave and
$s$-wave components of the OP.  A $d+is$ pairing state was also
found at the $\{110\}$-oriented surface or interface of a $d$-wave
superconductor in early work~\cite{Multi2}. A crucial difference
between the twin boundary of Fig.~\ref{FIG:OP-TB} and the
$\{110\}$-oriented surface of earlier work is that the dominant
$d$-wave component reaches a maximum at twin boundaries, whereas
it is strongly suppressed at the $\{110\}$-oriented surface or
interface. Experimentally, the existence of a
time-reversal-symmetry-breaking pairing state in high-$T_{c}$
cuprates is the subject of current debate~\cite{Multi3}.

\begin{figure}
\centerline{\psfig{figure=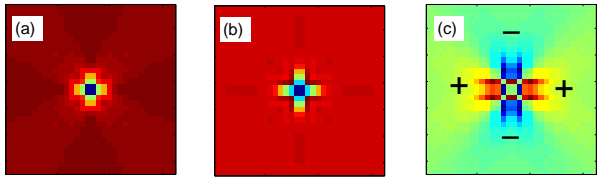,width=8.0cm,angle=0}}
\caption{Spatial variation of the SC OP for a single defect
displayed in Fig.~\ref{FIG:Lattice}(c)--- (a) The $s$-wave OP in
an $s$-wave superconductor, and (b) the $d$-wave and (c) extended
$s$-wave components of the OP in a $d$-wave superconductor. The
electron-lattice coupling constant $\alpha=3$. }
 \label{FIG:OP-DF3}
\end{figure}

As another example, we show in Fig.~\ref{FIG:OP-DF3} the spatial
variation of the superconducting OP around the single defect
(Fig.~\ref{FIG:Lattice}(c)) in both the $s$-wave and $d$-wave
superconductor cases. The OP is depressed at the center of the
defect, and reaches its defect-free bulk value at the length scale
$\xi_{0}$. Notice that for a lattice-deformation defect, which
affects the local electron hopping integral, the OP has a minimum
at four sites surrounding the defect center. It is different from
the case of an externally substituted unitary impurity, where the
minimum of OP is located only at the impurity site
itself~\cite{Zhu00}. The range of influence of such a defect can
be very large depending on the strength of electron-lattice
coupling--- the elasticity propagates the electronic response. The
$d$-wave energy gap has a sign change at the nodal directions of
the essentially cylindrical Fermi surface, but the $d$-wave OP
does not exhibit such a sign change in real space. When the defect
is introduced, an extended $s$-wave component of the OP is induced
when the dominant $d$-wave component is depressed at the defect.
Strikingly, the induced $s$-wave component has a sign change
across the diagonals of the square lattice, i.e.,
$\text{sgn}[\cos(2\theta)]$, where $\theta$ is the azimuthal angle
with respect to the crystalline $x$ axis. This is a direct
manifestation of the $d$-wave pairing symmetry in real space.
This feature can be understood from a phenomenological LG
free-energy density functional with a two-component SC OP:
$\mathcal{F}=\alpha_{s} \vert \Delta_{s}\vert^{2} +
\alpha_{d}\vert \Delta_{d} \vert^{2} +\beta_{1} \vert \Delta_{s}
\vert^{4} +\beta_{2} \vert \Delta_{d} \vert^{4} +\beta_{3} \vert
\Delta_{s} \vert^{2} \vert \Delta_{d} \vert^{2} +\beta_{4}
(\Delta_{s}^{*2} \Delta_{d}^{2} +\Delta_{d}^{*2}\Delta_{s}^{2})
+\gamma_{s} \vert \nabla \Delta_{s} \vert^{2} +\gamma_{d} \vert
\nabla \Delta_{d} \vert^{2} + \gamma_{sd} [\partial_{x} \Delta_{s}
\partial_{x} \Delta_{d}^{*} -\partial_{y} \Delta_{s}
\partial_{y} \Delta_{d}^{*} +\text{c.c}] $, where we take
$\alpha_{s}$ to be always positive while
$\alpha_{d}=\alpha_{d0}(T/T_{d0}-1)$ such that there exists only a
single transition into a $d$-wave pairing state in a homogeneous
system. When a defect is introduced, the $d$-wave component is
depressed. Since the $s$-wave component itself should be very
small, the term $\Delta_{s}^{*2} \Delta_{d}^{2}
+\Delta_{d}^{*2}\Delta_{s}^{2}$ is only a higher-order correction.
Therefore, it is the mixed-gradient term that induces the $s$-wave
component and also determines the relative phase to be $0$ or
$\pi$. The sign change is also evident by exchanging the $x$ and
$y$ components of the position coordinate in the mixed-gradient
term.

\begin{figure}
\centerline{\psfig{figure=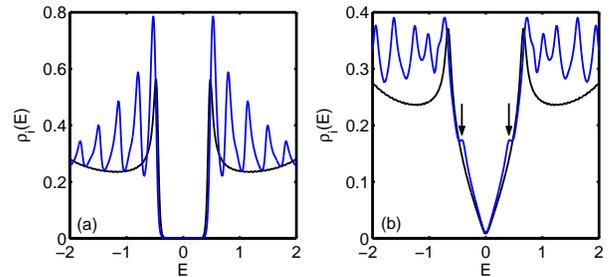,width=8.0cm,angle=0}}
\caption{The local density of states at a twin boundary in
$s$-wave (a) and $d$-wave superconductors. Also shown the LDOS
(black lines) for a single domain. The electron-lattice coupling
constant $\alpha=3$.} \label{FIG:LDOS-TB}
\end{figure}

Once the self-consistency for the order parameter is obtained, we
calculate the LDOS:
\begin{equation} \rho_{i}(E)= -\sum_{n}[ \vert
u_{i}^{n} \vert^{2} f^{\prime}(E-E_{n}) + \vert v_{i}^{n}
\vert^{2} f^{\prime}(E+E_{n})]\;,
\end{equation}
where $f^{\prime}(E)$ is the derivative of the Fermi distribution
function with respect to the energy. The LDOS determines the
differential tunneling conductance, measurable by STM
experiments~\cite{Pan00}.
Figure~\ref{FIG:LDOS-TB} shows the LDOS at a domain wall for both
types of superconductors, where the modulation of the
superconducting OP forms a superlattice, with its maximum at the
domain wall playing the role of an off-diagonal potential barrier
($\Delta_{ij}$ in Eq.~(\ref{EQ:BdG})). For an $s$-wave
superconductor, the quasiparticles are gapped away with their
energy below the minimum SC OP. Outside the minimum of the pair
potential, energy bands are formed by the quasiparticle scattering
off the off-diagonal energy barriers at the domain walls. Except
for the gap about the Fermi energy ($E=0$), this is reminiscent of
the electronic structure in semiconductor superlattices (e.g.,
alternating GaAs/GaAlAs layers). Interestingly, the bottom of the
oscillation pattern follows the LDOS (black line) of a system
formed by a single rectangular domain. Similar oscillations are
obtained for the $d$-wave superconductor. However, the bottom of
the oscillations do not follow the single domain DOS (black line).
In addition, weak subgap peaks (labeled by arrows in
Fig.~\ref{FIG:LDOS-TB}(b)) appear symmetrically in the LDOS on the
domain wall but are absent in the single-domain LDOS. We speculate
that  these resonant states  are due to the gradient of the
$s$-wave gap component induced inside the domain.

\begin{figure}
\centerline{\psfig{figure=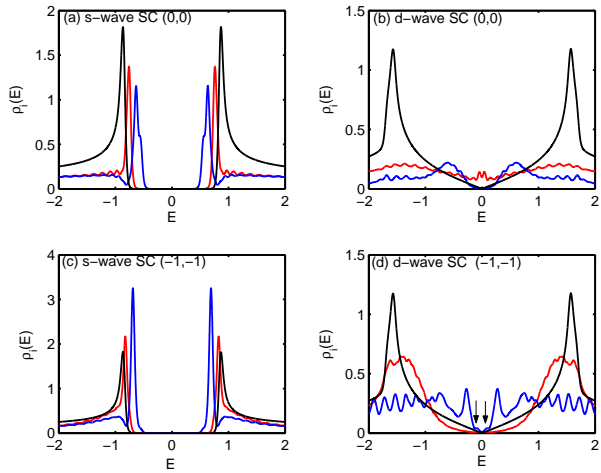,width=8.0cm,angle=0}}
\caption{The local density of states near the center of a defect
in $s$-wave (left column) and $d$-wave superconductors (right
column). The distance of the measured point away from the defect
is labeled by its coordinate. The electron-lattice coupling
constants are $\alpha=3$ (red lines) and $10$ (blue lines). Also
shown is the defect-free LDOS (black lines). }
\label{FIG:LDOS-DF3}
\end{figure}

As shown in Fig.~\ref{FIG:LDOS-DF3}, we have also calculated the
LDOS near the center of a single defect. The depression of the SC
OP at the defect makes a quantum-well-like profile of the energy
gap. The size and depth of the well is determined by the
electron-lattice coupling constant. Because of the difference
between the $s$-wave and $d$-wave pairings, the $s$-wave potential
well is closed everywhere, whereas the $d$-wave well has four
slits along the diagonals of the square lattice. In the $s$-wave
superconductor, the well is shallow and small for weak coupling,
which cannot trap low-lying quasiparticle bound states; for strong
coupling constants, the well is deep and large so that subgap
quasiparticle bound states are induced (the red and blue lines of
Fig.~\ref{FIG:LDOS-DF3}(a) and (c)). The energy of these low-lying
states must be inbetween the bottom and edge of the well.
Therefore, it is notable that the energy of these subgap states is
shifted toward the Fermi surface as the electron-lattice coupling
is increased (the blue line in Fig.~\ref{FIG:LDOS-DF3}(a) and
(c)). The situation here is also different from an $s$-wave vortex
core, where the OP at the core center must vanish such that the
low-lying bound states are always trapped~\cite{Multi4}. The
electronic structure at the defect in a $d$-wave superconductor
becomes even richer: For $\alpha=3$ (weak coupling as compared to
the band width of the non-deformed square lattice), the lattice
distortion plays the role of a weak defect for the quasiparticle
scattering.
In this case, a resonant peak with a dip exactly at the Fermi
energy is seen (the red line in Fig.~\ref{FIG:LDOS-DF3}(b)). The
overall peak comes from the scattering of quasiparticles off the
single-particle off-diagonal potential (i.e., local change of the
hopping integral as a response to the lattice deformation). This
lattice-deformation induced resonance state also exhibits Friedel
oscillations. Typically, the peak structure appears in the LDOS at
(0,0) (We label the four sites surrounding the defect center by
(0,0), (1,0), (1,1), (0,1)) and (-2,-2). In contrast to the case
of an extrinsic on-site potential-scattering
impurity~\cite{Balatsky95}, the LDOS spectrum is symmetric since
even the local particle-hole symmetry is preserved here.
For $\alpha=10$ (strong coupling), the $d$-wave OP is almost fully
depressed (less than 0.03), while the maximum of the induced
$s$-wave OP reaches 0.085. The local off-diagonal potential
becomes more finite ranged, which causes a local change of the
band width. The `resonant' peaks are pushed to higher energies
($\simeq \pm 0.3$) (the blue line of Fig.~\ref{FIG:LDOS-DF3}(d)).
Furthermore, small shoulders appear close to the Fermi energy (the
blue lines of Fig.~\ref{FIG:LDOS-DF3}(b) and (d)), which are
precursors of new Andreev resonance states. We have also computed
the LDOS without imposing self-consistency on the OP and found
that the double-peak structure is V-shaped with no existence of
the shoulders. This leads us to speculate that the new Andreev
resonance states come from the confinement of the induced $s$-wave
OP. However, these states are still delocalized because the
quasiparticles can leak out of the well through the slits along
the diagonal directions where the induced $s$-wave component
vanishes. All these features are unique to an elastic defect in a
$d$-wave superconductor with short coherence length.

In conclusion, we studied the effects of elastic lattice
deformation on the nanoscale electronic structure in
superconductors. We have shown that the SC OP is depressed in the
regions where the lattice deformation exists. The calculated LDOS
suggests that the electronic structure is strongly modulated in
response to the lattice deformation. In particular, it is possible
to trap low-lying quasiparticle states around the defects. Images
of these states will manifest the underlying long-range
anisotropic lattice deformation. These predictions can be directly
tested by STM experiments in new functional superconducting
materials. Our approach is readily extended to other elastic
textures and SC symmetries. Self-consistent coupling of the
elastic and SC textures on an equal footing will be pursued
elsewhere. It would also be interesting to study the electronic
response by using Monte Carlo generated microstructures as input,
which constitutes future work.

We thank  A. Saxena  and S. R. Shenoy for useful discussions. This
work was supported by the US Department of Energy through the Los
Alamos National Laboratory.

\end{document}